\documentclass[preprint,showpacs,showkeys,onecolumn,nofootinbib]{revtex4}
\pdfoutput=1
\usepackage[colorlinks=true,linkcolor=blue,urlcolor=blue,filecolor=black,citecolor=red,pdfstartview=FitV,pdftitle={},pdfsubject={},pdfkeywords={},pdfpagemode=None,bookmarksopen=true]{hyperref}
\usepackage{graphicx}
\usepackage{amsmath}
\usepackage{amsfonts}
\usepackage{amssymb,ulem}
\usepackage{color}%
\usepackage{dcolumn}
\usepackage{MnSymbol,wasysym}
\usepackage{braket}

\usepackage{eurosym}
\usepackage{calrsfs}
\usepackage[usenames,dvipsnames,svgnames]{xcolor}
\usepackage{color}%

\begin{document}


\title{Holographic heat engine in  Horndeski model with the $k$-essence sector}
\author{Shi-Qian Hu}
\email{mx120170256@yzu.edu.cn}
\author{Xiao-Mei Kuang}
\email{xmeikuang@yzu.edu.cn}
\affiliation{Center for Gravitation and Cosmology, College of Physical Science and Technology, Yangzhou University, Yangzhou, 225009, China}

\begin{abstract}
We study the extended thermodynamical properties of the charged black hole in Horndeski model with the $k$-essence sector.
Then we define a holographic  heat engine via the black hole.  We compute the engine efficiency in the large  temperature limit and
compare the results with the exact ones. With the given
specified parameters in the rectangular engine, the higher order  coupling suppresses the engine efficiencies.
\end{abstract}
\pacs{ 04.20.Gz, 04.20.-q, 03.65.-w}
\keywords{AdS black hole; Thermodynamics of black hole; Holographic heat engine}

\maketitle

\section{Introduction}
 Thermodynamics of the anti de Sitter (AdS) black holes are important fields for us to understand the nature of quantum gravity and it has been attracting much attention in the development of holographic gauge/gravity duality \cite{Maldacena:1997re,Gubser:1998bc,Witten:1998qj}. Recently, one of the most significant progress in the study of black hole thermodynamic is  extending the  study into a more general case by treating the cosmological constant, $\Lambda$, as the pressure of the thermal black hole system\cite{Kastor:2009wy,Dolan:2010ha,Cvetic:2010jb,Dolan:2011xt}. In this synopsis,  the thermodynamical volume is defined as the conjugation of the pressure or $\Lambda$, and it usually satisfies the Reverse Isoperimetric Inequality\cite{Cvetic:2010jb,Altamirano:2014tva} with a counterexample  proposed in
 \cite{Hennigar:2014cfa}. And the standard smarr formula also has to be modified such that  the mass of the black hole plays the role of the enthalpy of the thermodynamical system\cite{Kastor:2009wy}.
Phenomena in the extended thermodynamic of black hole including  Van der Waals like phase transition, triple points and reentrant phase transition and so on have been investigated. Readers can refer to the review paper \cite{Kubiznak:2016qmn} and therein for references.

Besides, based on the extended thermodynamic of AdS black hole, the author of \cite{Johnson:2014yja} proposed to define a traditional heat engine via an AdS black hole, which is realized by a circle in the pressure-volume phase space of the black hole. The defined engine is also called  holographic heat engine because the engine  circle represents a process defined on the space of the  dual field theory living in one dimension lower than the bulk.  In this engine, the input of heat, the exhaust of heat  and the mechanical work all  can be determined  from the gravitational system, so the engine efficient can directly be evaluated in the bulk.   The studies of holographic engine were soon extended in other modified gravitational model, such as  with Gauss-Bonnet correction\cite{Johnson:2015ekr}, in Born-Infeld corrected black hole\cite{Johnson:2015fva}, in rotational black hole\cite{Hennigar:2017apu,Johnson:2017ood}, in three dimensional black hole\cite{Mo:2017nhw} and so on\cite{Liu:2017baz,Wei:2016hkm,Mo:2018hav,Xu:2017ahm,Zhang:2018vqs}, and  many remarkable properties of engine efficiency  were observed. More recently, it was addressed in \cite{Johnson:2018amj} that the holographic heat engines defined via AdS black hole can be seen as a working substance correspond to specific combinations of conform field theory flows and deformations.

Since our world is far from being ideal, so it is more realistic to study the heat engine defined via black holes with momentum relaxation.  The engine efficiency modified by mass of graviton, which breaks the diffimophism system in the bulk and so  introduces the momentum relaxation in the dual theory\cite{Davison:2013jba},   has been carefully studied in \cite{Mo:2017nes,Hendi:2017bys}. One of us studied  the heat engine in the Einstein-Maxwell-Axions theory\cite{Fang:2017nse}, where the momentum relaxation is introduced by the linear massless axion fields\cite{Andrade:2013gsa}.

The aim of this paper is to continue to construct holographic heat engine with momentum relaxation.  We will work in the Horndeski gravity  with the $k$-essence sector proposed in \cite{Cisterna:2017jmv}.
In this theory, besides the minimal coupling of axions term, $X=\frac12\nabla^{\mu}\phi_{i}\nabla_{\mu}\phi_{i}$, the authors included  nonlinear terms in the form of an arbitrary power of $X$ and got  the exact black hole solution. So this paper can be treated as the extension of our previous work \cite{Fang:2017nse}.
Note that with the use of holography, it is  found in \cite{Cisterna:2017jmv} that the nonlinear terms modifies the dc conductivities. So we also expect that the terms will enrich the properties of the dual heat engine.

Following is the plan of this paper. We briefly review the AdS black hole solution in Horndeski model with the $k$-essence sector in section \ref{sec:review} and then study the extended thermodynamics in section \ref{sec:E-thermo}. In section \ref{sec:engine}, we define the holographic engine and get its general efficiency while in section \ref{sec:efficiency}, we compute  the engine efficiency in large temperature limit and then compare the results with  the exact results. Section \ref{sec:conclusion} is our conclusions and discussions. In this paper, we will work in the units with $G=c=\hbar=k_B=1$.
\section{Horndeski model with the $k$-essence sector}\label{sec:review}
We shall briefly review the four-dimensional Horndeski gravity  with the $k$-essence sector proposed in \cite{Cisterna:2017jmv}. The action was given  by
\begin{equation}
S=\frac{1}{16\pi}
\int\sqrt{-g}\left(R-2\Lambda-\frac{1}{4}F_{\mu\nu}F^{\mu\nu}-\sum_{i=1}^{2}(X_i+\gamma X_i^k)\right)d^4x\,,
\label{action}
\end{equation}
where  the cosmological constant is  $\Lambda=-3\ell^{-2}$ with $\ell$ the AdS radius and $X_{i}=\frac12\nabla^{\mu}\phi_{i}\nabla_{\mu}\phi_{i}$ with $i=1,2$. $\phi_i$ are massless scalar field and $F_{\mu\nu}$ is the field strength of Maxwell field.  The above action
 goes back to that for  the minimally coupled Einstein-Maxwell-axions gravity  studied in \cite{Bardoux:2012aw,Andrade:2013gsa}.  The equations of motions derived from the action are
the Klein-Gordon equation
\begin{equation}
\left((1+\gamma kX_i^{k-1})g^{\mu\nu}+\gamma
k(k-1)X_i^{k-2}\nabla^{\mu}\phi_i\nabla^{\nu}\phi_i\right)\nabla_{\mu}\nabla_{\nu}\phi_i=0,
\label{KG}
\end{equation}
the Maxwell equation
\begin{equation}
\nabla_{\mu}F^{\mu\nu}=0,
\label{Maxwell}
\end{equation}
and the Einstein equation
\begin{eqnarray}
G_{\mu\nu}+\Lambda g_{\mu\nu}=\frac{1}{2}\sum_i\left(\partial_{\mu}\phi_i\partial_{\nu}\phi_i-g_{\mu\nu}X_i+\gamma(kX_i^{k-1}\partial_{\mu}\phi_i\partial_{\nu}\phi_i-g_{\mu\nu}X_i^k)
-\frac{g_{\mu\nu}}{4}F^{\rho\sigma}F_{\rho\sigma}+F_{\mu\rho}F_{\nu}^{\rho}\right).
\label{EOM}
\end{eqnarray}
The above equations of motion admit the following exact black hole solution
\begin{eqnarray}
ds^2&=&-f(r)dt^2+\frac{dr^2}{f(r)}+r^2(dx_1^2+dx_2^2)\,\label{metriccharged}\\ \mathrm{with}~~~~
f(r)&=&\frac{r^2}{\ell^2}-\frac{2m}{r}-\frac{\lambda^2}{2}+\gamma\frac{\lambda^{2k}}{2^k(2k-3)}r^{2(1-k)}+\frac{q^2}{4
r^2}, \label{metricf}
\end{eqnarray}
and the matter fields
\begin{eqnarray}
\phi_1&=&\lambda x_1,\ \phi_2=\lambda x_2\ , \label{axions}\\
A&=&\left(\mu-\frac{q}{r}\right)dt\,,\label{At}
\end{eqnarray}
where $m,q$ are integral constants which are connected with the physical quantities of the black hole and the integral constant  $\mu$ is to guarantee the regular condition of Maxwell field at the horizon. So this solution is not exactly the one presented in  \cite{Cisterna:2017jmv} where they had the Coulomb form $A_t=-\frac{q}{r}$.
We note that as pointed out in \cite{Cisterna:2017jmv},  the coupling parameter $\gamma$ is required  to be positive to avoid phantom
contributions. And the null energy condition, finite ADM mass and asymptotically matching the GR black hole solution requires the restriction $k>3/2$ in the solution.

\section{The extended thermodynamics }\label{sec:E-thermo}
 The usual thermodynamical analysis has been achieved via the Euclidean approach in \cite{Cisterna:2017jmv}. Here, we will analyze the extended thermodynamical properties of the black hole solution \eqref{metriccharged}-\eqref{At}. To this end,  we connect the cosmological constant and the pressure of the system via \cite{Kastor:2009wy,Dolan:2010ha,Cvetic:2010jb,Dolan:2011xt}
\begin{equation}\label{defP}
P=-\frac{\Lambda}{8\pi}=\frac{3}{8\pi\ell^2},
\end{equation}
where the thermodynamic quantity conjugate to the pressure is defined as the thermodynamic volume of black holes.

Defining the location of horizon, $r_h$, which satisfies  $f(r_h)=0$ in \eqref{metricf}, the integral constant $m$ is solved as
\begin{equation}
m=\frac{\gamma  2^{-k-1} \lambda ^{2 k} r_h^{2 (1-k)+1}}{2 k-3}+\frac{4}{3} \pi  P r_h^3+\frac{q^2}{8 r_h}-\frac{\lambda ^2 r_h}{4},
\end{equation}
where we have used \eqref{defP}. The Hawking temperature of the  black hole is given by,
\begin{equation}\label{Temp}
T =\frac{f'(r)\mid_{r\to r_h}}{4\pi}=2 P r_h-\frac{q^2}{16 \pi  r_h^3}-\frac{\lambda ^2}{8 \pi  r_h}-\frac{\gamma  2^{-k-2} \lambda ^{2 k} r_h^{1-2 k}}{\pi },
\end{equation}
and the entropy is
\begin{equation}\label{eq-S}
S =\frac{\mathcal{V}_2 r_h^2}{4}.
\end{equation}
The mass and charge of the black hole are connected with the parameters as\cite{BHMASS}
\begin{equation}\label{MQ}
M=\frac{4m\mathcal{V}_2}{16\pi}=\frac{\mathcal{V}_2}{16\pi}\left(\frac{\gamma  2^{1-k} \lambda ^{2 k} r_h^{2 (1-k)+1}}{2 k-3}+\frac{16}{3} \pi  P r_h^3+\frac{q^2}{2 r_h}-\lambda ^2 r_h\right),
~~Q=\frac{\mathcal{V}_2}{16\pi}q
\end{equation}
where $\mathcal{V}_2$ is the volume of the two dimensional flat space. Subsequently, the thermodynamical volume as the conjugation of the pressure  is
\begin{equation}\label{eq-V}
V=\left(\frac{\partial M}{\partial P}\right)_{S,Q}= \frac{\mathcal{V}_2}{3}r_h^3
\end{equation}
and the electric potential of the black hole is\cite{Caldarelli:1999xj,Cvetic:1999ne}
\begin{equation}
\Phi =A_{t}\chi^{\mu}\mid_{r\to\infty}-A_{t}\chi^{\mu}\mid_{r=r_h}=\mu=\frac{q}{r_h}=\left(\frac{\partial M}{\partial Q}\right)_{S,P}.
\end{equation}
It is straightforward to verify that the first law of thermodynamics
\begin{equation}\label{eq-dM0}
d M=T dS+\Phi dQ+V dP
\end{equation}
is satisfied.

According to the dimensional analysis\cite{Kastor:2009wy}, we obtain that the modified Smarr relation for the black hole is\footnote{We thank Yen Chin Ong for the helpful discussion on the Smarr formula.}
\begin{equation}\label{eq-Smarr}
M=2TS+\Phi Q-2PV+(2k-2)\varphi \gamma,
\end{equation}
where $\varphi$ is the conjugation of $\gamma$ with dimension $2k-2$
 \begin{equation}
 \varphi=\left(\frac{\partial M}{\partial \gamma}\right)_{S,P,Q}=\frac{ \mathcal{V}_2 \lambda ^{2 k} r_h^{3-2k}}{ (2 k-3)2^{k+3}\pi}.
 \end{equation}
It is obvious that without the higher order coupling, i.e, $\gamma=0$ or $k=1$, the Smarr formula is standard which goes back to the result found in \cite{Fang:2017nse}.
The above Smarr like relation leads us to consider  the role of the mass from internal energy to enthalpy, and so  the first law of black hole
thermodynamics \eqref{eq-dM0} should be modified as
\begin{equation}\label{eq-dM}
d M=dH=T dS+\Phi dQ+V dP+\varphi d\gamma.
\end{equation}

From the expression of temperature, we get the state equation
\begin{equation}\label{stateE}
P =\frac{T}{2 r_h}+\frac{q^2}{32\pi  r_h^4}+\frac{\lambda ^2}{16 \pi  r_h^2}+\frac{\gamma  2^{-k-3} \lambda ^{2 k} r_h^{-2 k}}{\pi }.
\end{equation}
In the right hand side of \eqref{stateE}, all terms are positive since $\gamma$ can only be positive, so we can not observe the $P-V$ criticality in the extended phase space proposed in \cite{Chamblin:1999tk,Chamblin:1999hg,Kubiznak:2012wp}. This is not necessary to construct dual heat engine to the black hole, which we will study in next sections.

\section{Holographic heat engine and its general efficiency}\label{sec:engine}

We will study the holographic heat engine built via the black hole solution described in previous section.
Before the construction, we have to exact two important thermodynamical physical quantities, the specific heat with the constant volume, $C_V$, and the specific heat with constant pressure, $C_P$, of the black hole. The general specific heat is defined as  $C=T(\partial S/\partial T)$, so we can treat both $T$ and $S$ as functions of the horizon $r_h$ to achieve the calculation.

Differentiation of \eqref{eq-S} gives us
\begin{equation}
\frac{\partial S}{\partial r_h}=\frac{\mathcal{V}_2}{2}r_h
\end{equation}
The differentiation to the expression of temperature \eqref{Temp} is
\begin{equation}
dT= \left(2 P+\frac{3 q^2}{16 \pi  r_h^4}+\frac{\lambda ^2}{8 \pi  r_h^2}+\frac{\gamma  2^{-k-2} (2 k-1) \lambda ^{2 k} r_h^{-2 k}}{\pi }\right)d r_h+2  r_h d P
\end{equation}
from which we obtain
\begin{equation}
\frac{\partial T}{\partial r_h}=\frac{2 P+\frac{3 q^2}{16 \pi  r_h^4}+\frac{\lambda ^2}{8 \pi  r_h^2}+\frac{\gamma  2^{-k-2} (2 k-1) \lambda ^{2 k} r_h^{-2 k}}{\pi }}{1-2  r_h \frac{\partial P}{\partial T}}.
\end{equation}
Then the general formula of the specific heat is
 \begin{eqnarray}\label{eq-C}
 C&=&T\frac{\partial S}{\partial T}=T\frac{\left(\frac{\partial S}{\partial r_h}\right)}{\left(\frac{\partial T}{\partial r_h}\right)}=\left(1-2  r_h \frac{\partial P}{\partial T}\right)\frac{2 \pi  P r_h^6-\frac{1}{16} q^2 r_h^2-\frac{1}{8} \lambda ^2 r_h^4-\gamma 2^{-k-2} \lambda ^{2 k} r_h^{6-2 k}}{4
   \pi  P r_h^4+\frac{1}{4} \lambda ^2 r_h^2+\frac{3 q^2}{8}+\gamma  2^{-k-1} (2 k-1) \lambda ^{2 k} r_h^{4-2 k}}.
 \end{eqnarray}

In the case with constant volume which means also constant $r_h$ via \eqref{eq-V}, the state equation~(\ref{stateE}) gives us
$(\partial P/\partial T)_{V}=\frac{1}{2r_h}$. Thus, the specific heat at constant volume is reduced to
 \begin{equation}
 C_{V}=T\frac{\partial S}{\partial T}\big|_{V}=0.
 \end{equation}
which implies that adiabats and isochores are equivalent. We will see soon that this is a helpful property in the construction of
the cycle of the holographic heat engine.

In the case with constant pressure, we have $\partial P/\partial T=0$, so the specific heat at constant pressure, $C_P$, is
  \begin{eqnarray}\label{eq-Cp}
 C_P&=&T\frac{\partial S}{\partial T}\big|_P=\frac{2 \pi  P r_h^6-\frac{1}{16} q^2 r_h^2-\frac{1}{8} \lambda ^2 r_h^4-\gamma 2^{-k-2} \lambda ^{2 k} r_h^{6-2 k}}{4
   \pi  P r_h^4+\frac{1}{4} \lambda ^2 r_h^2+\frac{3 q^2}{8}+\gamma  2^{-k-1} (2 k-1) \lambda ^{2 k} r_h^{4-2 k}}.
 \end{eqnarray}
\begin{figure}[h]
{\centering
\includegraphics[width=3in]{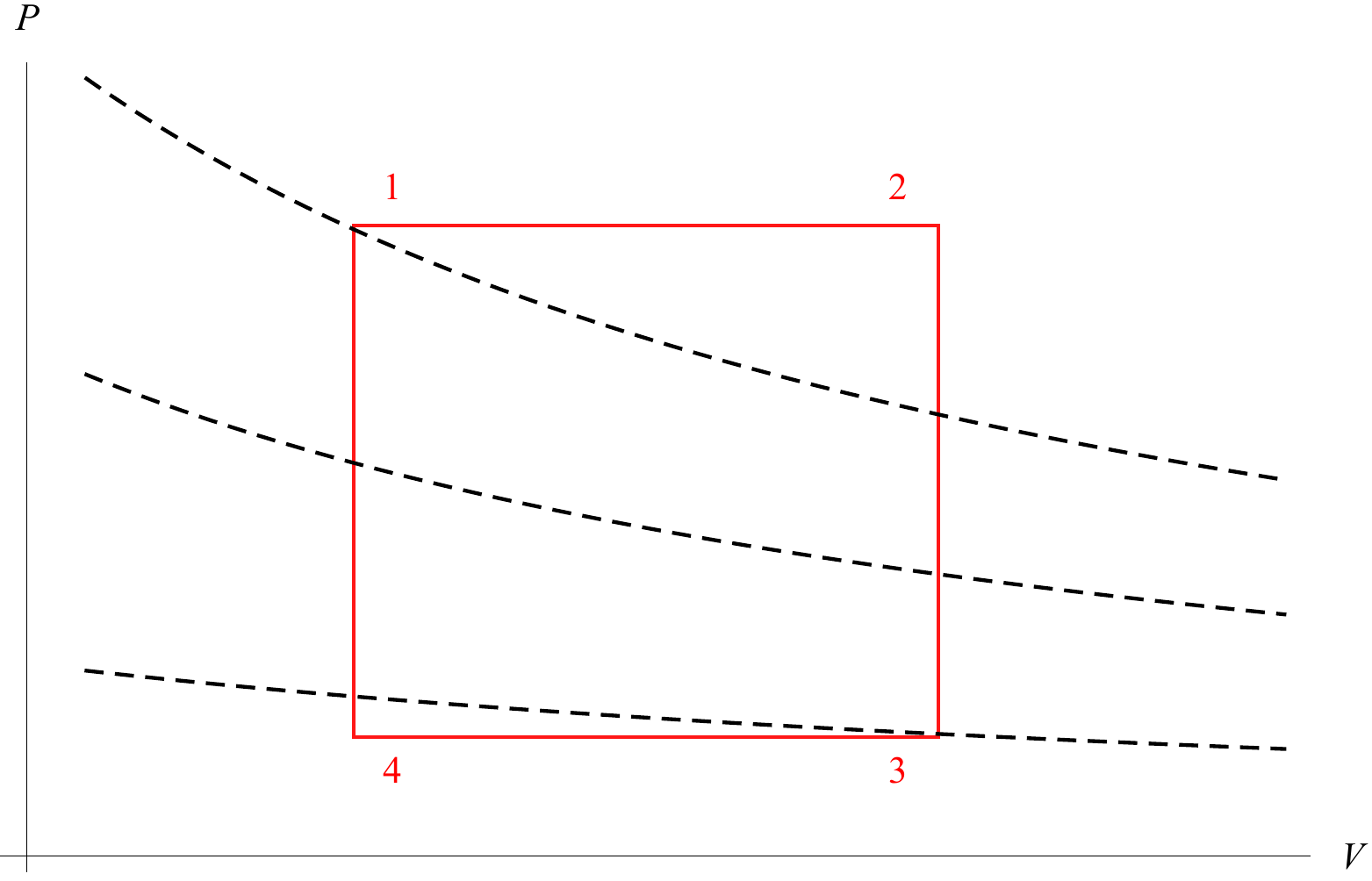}
   \caption{Cartoon of the engine.}   \label{fig-engine}}
\end{figure}

Now, we are ready to define a heat engine dual to the black hole. Following \cite{Johnson:2015fva,Hennigar:2017apu,Xu:2017ahm,Johnson:2017ood,Mo:2017nhw,Liu:2017baz,Wei:2016hkm},we consider a rectangle cycle in the $P-V$ plane. The cycle is consisted by two isobars and two isochores as shown in figure \ref{fig-engine} where  $1,2,3,4$ are four corners in the thermal flow cycle.  We will  express the relevant quantities evaluated with the use the subscripts $1, 2, 3, 4$  at the related corners. The engine efficiency is computed by
\begin{equation}
\label{eq-eta}\eta=\frac{W}{\mathcal{Q}_H}=\frac{(V_2-V_1)(P_1-P_4)}{\int_{T_1}^{T_2} C_P(P_1,T) dT}
\end{equation}
where  $W$ is the work done by the engine and $\mathcal{Q}_H$ is the input of the heat
due to the equivalence of adiabats and isochores in the circle, respectively.
It is noticed  that in $P-V$ plane, the isotherms at temperatures $T_h$ and $T_l$ with $T_h>T_l$ give the Carnot efficiency $\eta_C$ and for our engine,  it is
\begin{equation}
\eta_C=1-\frac{T_l}{T_h}=1-\frac{T_4}{T_2}.
\end{equation}

The effect of the momentum  $\lambda$ on the engine efficiency has been studied by one of us in \cite{Fang:2017nse}. So here we will mainly study the engine efficiency  $\eta$ modified by the higher terms of the axions field . We will set $q=0.1$ and the volume $\mathcal{V}_{2}=1$ without loss of generality.

\section{Engine efficiency in large temperature limit}\label{sec:efficiency}
To evaluate the efficiency, in general, we can cancel $r_h$ in \eqref{eq-Cp} using \eqref{Temp} to rewrite  $C_P$ as a function of $(P,T)$ and then  applying the definition \eqref{eq-eta}. However, the integration  is difficult to proceed. So in order to study the efficiency directly from the definition. We consider the large temperature limit, i.e., $T \gg \lambda,q,\gamma$, which means that $1/T$ can be treated as a small quantity. Solving $r_h$ in term of large $T$ from \eqref{Temp}, we obtain
\begin{eqnarray}
r_h&=&\frac{T}{2 P}+\frac{\lambda ^2}{8 \pi  T}+\frac{8 \pi  P^2q^2-\lambda ^4 P}{32 \pi ^2 T^3}+\frac{\lambda ^6 P^2-16 \pi  \lambda ^2 P^3 q^2}{64 \pi ^3 T^5}+\cdots\notag\\
&+&\gamma\left(\frac{  2^{k-4} \lambda ^{2 k} P^{2 k-2} }{\pi T^{2 k-1 }}-\frac{ 2^{k-5} k \lambda ^{2 k+2} P^{2 k-1} }{\pi ^2T^{2 k+1}}+\cdots\right).
\end{eqnarray}
Then from \eqref{eq-V} and \eqref{eq-Cp}, we get the thermodynamic volume and the specific heat at constant pressure in large $T$ limit as
\begin{eqnarray}
V&=&\frac{T^3}{24 P^3}+\frac{\lambda ^2 T}{32 \pi  P^2}+\frac{q^2}{16 \pi  T}+\frac{\lambda ^6-48 \pi  \lambda ^2 P q^2}{1536 \pi ^3 T^3}-\frac{128 \pi ^2 P^3 q^4-48 \pi  \lambda^4 P^2 q^2+\lambda ^8 P}{2048 \pi ^4 T^5}+\cdots\notag\\
   &+&\frac{\gamma}{4 \pi}  \left(\frac{2^{k-5} (1-k) \lambda ^{2 k+2} P^{2 k-3}}{ T^{2 k-1}}+\frac{2^{k-4} \lambda ^{2 k} P^{2 k-4}}{ T^{2 k-3}}+\cdots\right),
\end{eqnarray}
\begin{eqnarray}
C_P&=&\frac{T^2}{8 P^2}+\frac{\lambda^4-16 \pi  P q^2}{128 \pi ^2 T^2}+\frac{24 \pi  \lambda ^2 P^2 q^2-\lambda ^6 P}{128 \pi ^3 T^4}+\frac{960 \pi ^2 P^4 q^4-480 \pi  \lambda ^4 P^3 q^2+15 \lambda ^8 P^2}{2048 \pi ^4 T^6}+\cdots\notag\\&+&\gamma  \left(\frac{2^{k-5} (k-1) \lambda ^{2 k} P^{2 k-3} }{\pi T^{2 k-2} }+\frac{2^{k-7} k (2 k-1) \lambda ^{2 k+2} P^{2 k-2} }{\pi ^2T^{2 k}}+\cdots\right),
\label{eq:4dexpansions}
\end{eqnarray}
respectively.

Subsequently, we can substitute the above expression of $V$ and $C_P$ into \eqref{eq-eta} to calculate the  efficiency of the engine. Considering the relations of each corner  in the circle, we find $\eta$ is  finally determined by the values of $(P_1,P_4, T_1,T_2)$ and we will not write  down the detailed expression due to the complexity.

\begin{figure}[h]
{\centering
\includegraphics[width=2.5in]{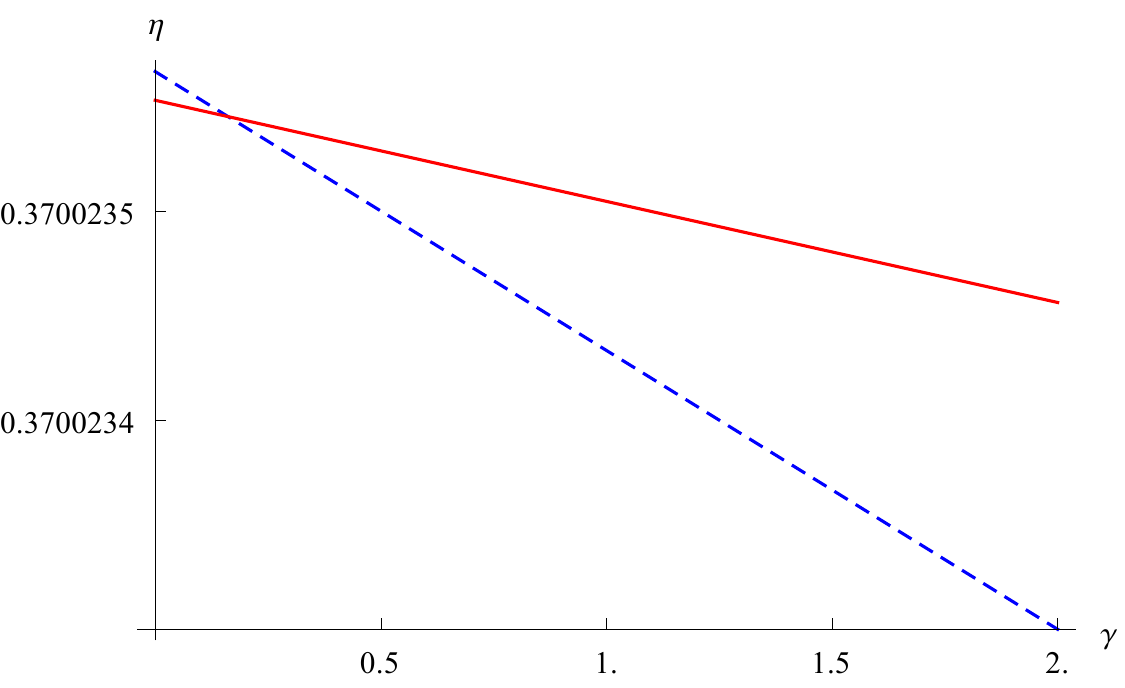}\hspace{0.15cm}
\includegraphics[width=2.5in]{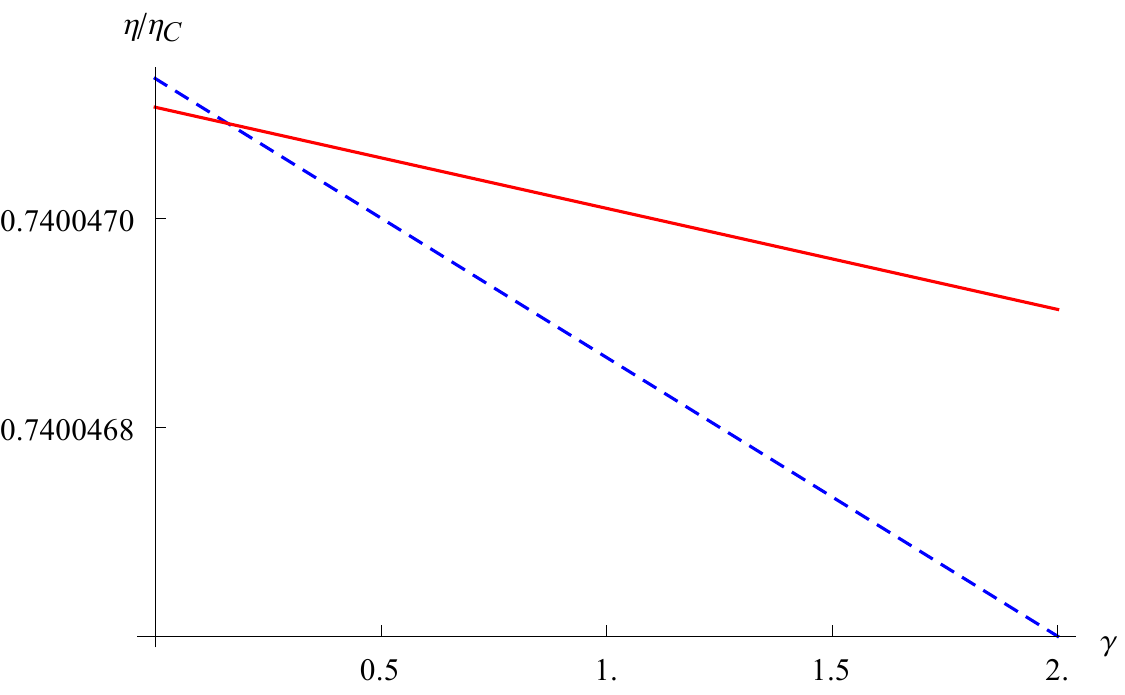}
   \caption{Engine efficiency effected by $\gamma$. We set $k=2$. }   \label{fig-eta-g2}}
\end{figure}

 We now study the affection of higher terms of axions on $\eta$. It was addressed in \cite{Johnson:2014yja} that one can process by choose different schemes on the given specified quantities. Here, we will work with given specified $(T_2,T_4,V_2,V_4)$ because in this scheme the Carnot efficiency $\eta_C=T_4/T_2$ will not change with parameters. Recalling the state equation \eqref{stateE}, we can calculate  $P_1=P_2$ via $(T_2,V_2)$ and $P_4$ via  and  $(T_4,V_4)$, respectively. Then we can further calculate $T_1$ via $(V_1=V_4,P_1)$. The efficiencies $\eta$ and $\eta/\eta_C$ change as $\gamma$  are shown by dashed lines  in figure \ref{fig-eta-g2} which shows that  $\eta$  and $\eta/\eta_C$ are both suppressed by increasing the coupling parameter. $\eta$ is always lower than the Carnot efficiency which is expected because  Carnot cycle is the most efficient. We also list the related efficiency for samples of exponent $k$ in table \ref{table1}. As $k$ increases, $\eta$ increases slowly, so does $\eta/\eta_C$.

\begin{widetext}
\begin{table}[h]
\begin{center}
\begin{tabular}{|c|c|c|}
 \hline
$k$&$\eta$ &$\eta/\eta_c $  \\  \hline
$2$&0.37002343 & 0.74004687  \\  \hline
$5/2$ &0.37002355 &0.74004712   \\ \hline
$3$&0.37002357 &0.74004713   \\ \hline
 \end{tabular}
 \caption{\label{table1}Engine efficiency effected by $k$ with $\gamma=1$.}
 \end{center}
\end{table}
\end{widetext}

In order to examine whether the above results in large $T$ limit are reliable, we  shall double
check them with the exact formula of efficiency
\begin{equation}\label{eq-etaA}
\eta=1-\frac{M_3-M_4}{M_2-M_1}
\end{equation}
proposed in \cite{Johnson:2016pfa} where $M_1,M_2,M_3,M_4$ denote the related mass of the black hole evaluated at each  corner in the cycle, i.e.,  the values of \eqref{MQ}
computed at each corner. With the same setup, the results of $\eta$ and $\eta_c$ evaluated by \eqref{eq-etaA} are shown by solid lines in figure \ref{fig-eta-g2}. Comparing with the dashed and solid lines in the plots, though the results in large $T$ limit have derivation from the exact results, but the rules are the same. The agreement is also fulfilled for different power exponent $k$.

\section{Conclusion}\label{sec:conclusion}
In this paper, we focused on the charged AdS black hole in Horndeski gravity  with the $k$-essence sector proposed in \cite{Cisterna:2017jmv} where higher order couplings of axions field were introduced.
We studied the  extended thermodynamics of the black hole and  derived  the modified  smarr formula as well as the generalized first law of thermodynamic. We found that the coupling parameter $\gamma$ should be treated as a thermodynamical quantity to achieve them.

Then based on the extended thermodynamics, we built the holographic  heat engine by the AdS black hole. We especially studied the effect of higher order couplings  on the efficiency of heat engine in
large temperature limit. With the given specified $(T_2,T_4,V_2,V_4)$,   the efficiency is always lower than the Carnot efficiency as we expect, and it is suppressed by the stronger coupling parameter. Moreover, as the exponent of higher coupling increases, the efficiency is enhanced slowly. Finally, we also calculated the exact efficiencies by the method proposed \cite{Johnson:2016pfa} and it was shown that our results in large temperature somehow is reliable. It is notable that it is also interesting to study the heat engine with momentum relaxation constructed  via black string or p-brane exactly solved in \cite{Cisterna:2017qrb}.

In this paper, we have focused on the rectangular engine shown in figure \ref{fig-engine}, the studies on  how the momentum relaxations terms affect on the engine efficiency in circular engine  \cite{Chakraborty:2016ssb} and axially symmetric engine as  well as riangular type engine\cite{Rosso:2018acz}, can be further extended. As  claimed in \cite{Johnson:2018amj} that the holographic heat engines can be seen as a working substance correspond to specific combinations of conform field theory flows and deformations, so  it would be significant  to investigate from holography that the role of the higher couplings in the flows, which may help us understand the deep physic of the phenomena we  observed. We shall study this issue in the near future.

\section*{Acknowledgement}
This work is supported by the Natural Science Foundation of China under Grant No.11705161 and Natural Science Foundation of Jiangsu Province under Grant No.BK20170481.  Xiao-Mei Kuang appreciates  Li-Qing Fang for the related  collaborations and she also appreciates Adolfo Cisterna for nice correspondence. We thank Bo Liu and  group members in CGC of YZU (http://www.cgc-yzu.cn) for helpful discussions.

\end{document}